\documentclass[preprint,12pt,authoryear]{elsarticle}

\usepackage{amssymb}

\journal{New Astronomy}

\begin{document}

\begin{frontmatter}



\title{Modeling particle acceleration and non-thermal emission in
supernova remnants}


\author[ad1]{S. Orlando}
\ead{salvatore.orlando@inaf.it}

\author[ad2,ad1]{M. Miceli}
\author[ad1]{S. Ustamujic}
\author[ad2,ad3]{A. Tutone}
\author[ad2,ad1]{E. Greco}
\author[ad4]{O. Petruk}
\author[ad1]{F. Bocchino}
\author[ad2,ad1]{G. Peres}

\address[ad1]{INAF - Oss. Astronomico di Palermo, Piazza
del Parlamento 1, 90134 Palermo, Italy}
\address[ad2]{Dip. di Fis. e Chim., Universit\`a
di Palermo, Via Archirafi 36, 90123 Palermo, Italy}
\address[ad3]{INAF - IASF, Via Ugo La Malfa 153, 90146 Palermo, Italy}
\address[ad4]{Inst. for Appl. Probl. in Mech. and
Math., Naukova Street, 3-b Lviv 79060, Ukraine}

\begin{abstract}
According to the most popular model for the origin of cosmic rays
(CRs), supernova remnants (SNRs) are the site where CRs are
accelerated. Observations across the electromagnetic spectrum
support this picture through the detection of non-thermal emission
that is compatible with being synchrotron or inverse Compton radiation
from high energy electrons, or pion decay due to proton-proton
interactions. These observations of growing quantity and quality
promise to unveil many aspects of CRs acceleration and require more
and more accurate tools for their interpretation. Here, we show how
multi-dimensional MHD models of SNRs, including the effects on shock
dynamics due to back-reaction of accelerated CRs and the synthesis
of non-thermal emission, turned out to be very useful to investigate
the signatures of CRs acceleration and to put constraints on the
acceleration mechanism of high energy particles. These models have
been used to interpret accurately observations of SNRs in various
bands (radio, X-ray and $\gamma$-ray) and to extract from them key
information about CRs acceleration.
\end{abstract}

%
%


\end{frontmatter}


\section{Introduction}

Many lines of evidence suggest that supernova remnants (SNRs) are
the site where CRs diffusive shock acceleration (DSA) occurs. Roughly
10\% of the supernova (SN) energy would be required to be transmitted
to CRs to satisfy observations. In fact, observations in various
bands support this idea through the detection of non-thermal emission
that is compatible with being synchrotron radiation from CR electrons.
Conversely, the direct evidence of CR ions is difficult to find
because they do not radiate efficiently; some hints of their existence
come from the analysis of $\gamma$-rays.

On the other side, the effects of CR ions on the dynamics of the
shock waves (and, therefore, on the thermal emission from the shell of SNRs
in different wavelength bands) are often invoked as indirect evidence
for their existence. A possible signature of CR ions reported in
the literature is the separation between the forward shock (FS) and
the contact discontinuity (CD) in young SNRs; this separation should
be smaller than expected, if CRs acceleration is efficient (e.g.
\citealt{2010A&A...509L..10F}).
Furthermore CR ions affect the evolution of the magnetic field and
the signature of their presence can be found in the non-thermal
emission from the electrons in different bands (radio, X-rays,
$\gamma$-rays; e.g. \citealt{2014ApJ...789...49F}).

The numerical simulations involving CRs acceleration at shock fronts
of SNRs is studied by following two complementary approaches. The
first is based on highly detailed models implemented with
particle-in-cell (PIC) and hybrid (kinetic ions-fluid electrons)
simulations (e.g. \citealt{2008ApJ...682L...5S, 2014ApJ...783...91C,
2014ApJ...794...46C, 2020PrPNP.11103751P}). These models are based
on first principles and consider few or no parameters/approximations.
However, they require high computational resources and they have
limited dynamical ranges. For instance, these models cannot simulate
the whole remnant interacting with the inhomogeneous interstellar
medium (ISM) or describe the complete evolution from the SN event
to the full-fledged SNR.

The second approach is based on three-dimensional (3D)
(magneto)-hydrodynamic (HD/MHD) simulations coupled with semi-analytic
models which provide a general description of the CRs acceleration
(e.g. \citealt{2010A&A...509L..10F, 2012ApJ...749..156O}). This
approach is less accurate in the prescription of CRs than the
previous one but allows to simulate the evolution of the whole
remnant from the SN explosion to its interaction with the inhomogeneous
and magnetized ISM. These models include the microphysics relevant
for the evolution of a SNR (e.g. magnetic-field-oriented
thermal conduction\footnote{In the presence of an organized
magnetic field, the thermal conduction is highly anisotropic and
it can be extraordinarily reduced in the direction transverse to
the field. In this case, the models describe a magnetic-field-oriented
thermal conduction (e.g.  \citealt{2008ApJ...678..274O}).},
deviations from equilibrium of ionization and from electron-proton
temperature equilibration, radiative cooling, etc.) and they are
constrained by multi-$\lambda$ observations.

While the first approach (based on PIC, hybrid simulations) is
mainly oriented toward an accurate description and a deep physical
insight of the microscopic mechanisms for particle acceleration,
the second one (based on HD/MHD simulations coupled with semi-analytic
models of particle acceleration) is mostly oriented toward a
description of the macroscopic effects of particle acceleration on
the dynamics and evolution of SNRs, especially during the interaction
of the remnants with the inhomogeneous ISM. This latter approach
can be very powerful for the interpretation of observations, and
for obtaining information and constraints on the mechanisms of particle
acceleration (e.g. \citealt{2011A&A...531A.129B, 2016A&A...593A..26M}).

Here, we discuss some of the recent achievements obtained using the
second approach, the capabilities and limitations of the HD/MHD
models describing the macroscopic effects of particle acceleration,
and the methodologies generally adopted to study and use these
models for the interpretation of observations in different bands.
We limit the discussion to models mostly developed by our
group; so we warn the reader that the review is highly biased on
results we obtained in the last years.

\section{Modeling the macroscopic effects of particle acceleration
on the dynamics of SNRs}

Models of SNRs, which include the effects of particle acceleration,
have to describe all the features characterizing a remnant (multiple
shocks, HD instabilities developing during the evolution, interaction with
inhomogeneities of the ISM, etc.) and to synthesize the thermal and
non-thermal emission in various bands. Examples of these models are
those produced with the CR-hydro-NEI (ChN) code
(\citealt{2008ApJ...686..325L, 2012ApJ...750..156L}), a one-dimensional
(1D) code appropriate to describe young to middle-aged SNRs. It
includes nonlinear DSA physics, and describes the CR back-pressure,
the particle escape, and the magnetic turbulence generation.  The
output of the code (multi-$\lambda$ spectra, spatial variation of
emission, thermal and non-thermal emission, etc.) can be compared
directly with observations, thus providing a key for the interpretation
of the data (and also to get a feedback on the models).

Nevertheless, although very sophisticated and rich in physics, the
ChN code is 1D and, therefore, it is not adapt to describe the
asymmetries which develop during the remnant evolution and the
complex interaction of the SNR with the inhomogeneous ISM. In these latter
cases a full 3D description is required. \cite{2010A&A...509L..10F}
developed a 3D model of SNR coupled with a semi-analytical kinetic
model of shock acceleration (\citealt{2002APh....16..429B,
2004APh....21...45B}) by means of an effective adiabatic index which
depends on time. The non-linear model of DSA includes the escape
of particles with the highest energy upstream of the shock and the
effect of Alfv\'en wave heating in the precursor. With their model,
the authors have studied the time-dependent compression of the
region between the FS and the reverse shock (RS) due to the
back-reaction of accelerated particles, for different values of the
injection efficiency. They found that the density profiles and the
thickness of the mixing region depend critically on the injection
level of particles.

Moreover, this model provides a theoretical support to the idea
that a greater shock compression ratio and a thinner
shell of shocked ISM are direct consequences of the energy losses
to CRs at the FS. Evidence of thinner shell of shocked ISM have
been found in some SNRs (e.g. Tycho and SN~1006), suggesting the
presence of efficient CRs acceleration (e.g. \citealt{2001ApJ...560..244B,
2009A&A...501..239M, 2010A&A...509L..10F}). Some authors, however,
have raised doubts if the CRs energy losses at shock fronts alone can
naturally explain observations as those of Tycho and SN~1006 (e.g.
\citealt{2011ApJ...735L..21R, 2012ApJ...749..156O}).

In fact, the observations show that, in general, remnant outlines
are characterized by several protrusions by ejecta fragments. These
features may be explained as due to a thinner shell of shocked ISM
caused by energy losses to CRs (e.g. \citealt{2008ApJ...680.1180C,
2011ApJ...735L..21R}). However, the number of protrusions in regions
dominated by thermal emission is more or less the same as in regions
dominated by non-thermal emission and this evidence suggests that
the occurrence of protrusions cannot be explained in terms of energy
losses to CRs alone.  In addition, extreme energy losses are needed
to allow a significant fraction of the ejecta to approach or even
overtake the FS to form protrusions. For instance,
\cite{2001ApJ...549.1119W} have shown that an extreme effective
adiabatic index of the order of 1.1 would be necessary to produce
evident protrusions in a remnant outline. Furthermore, in the case
of SN~1006, the ratio of FS to CD radii is almost the same at all
azimuthal angles (\citealt{2009A&A...501..239M}), in regions dominated
by non-thermal emission (where CRs acceleration is efficient) and
in those dominated by thermal emission (where the CRs acceleration
efficiency should be negligible). This is in stark contrast with
the predictions that the ratio of FS to CD radii should be the
minimum in regions of efficient CRs acceleration.

All these lines of evidence suggest that FS-CD separation may be
misleading as indicator for the efficiency of CRs acceleration.
\cite{2011ApJ...735L..21R} suggested that the ejecta structure
of the explosion itself can play a role in determining the frequent
occurrence of protrusions (even in regions with no evident CRs
acceleration). In other words, the observed structure of the mixing
region between the FS and the RS may originate in the intrinsic
structure of the ejecta.

\section{Modeling the evolution from the SN to the SNR including
the CRs acceleration} 

The difficulties of models accounting only for the back-reaction
of accelerated CRs in the interpretation of observations have
stimulated the development of 3D HD/MHD models which describe
accurately the structure of the ejecta originating from the SN
explosion. These models describe the evolution from the SN to the
development of its full-fledged remnant (e.g. \citealt{2015ApJ...810..168O,
2016ApJ...822...22O, 2017ApJ...842...13W, 2019ApJ...877..136F,
2020ApJ...888..111O, 2020A&A...636A..22O, 2020A&A...642A..67T,
2020arXiv200901789O}).  The strategy is based on the coupling between
SN models and SNR models. The SN models provide the initial conditions
for the SNR simulations few hours after the SN event and, among
other things, include: the explosive nucleosynthesis through a
nuclear reaction network, the energy deposition due to radioactive
decays of isotopes synthesized in the SN, the gravitational effects
of the central compact object, and the fallback of material on the
compact object (e.g. \citealt{2017ApJ...842...13W, 2020ApJ...888..111O}).
The pre-SN structure of the stellar progenitor is derived by available
stellar evolution codes. The SNR models include: the effects of an
ambient magnetic field, the radiative cooling, the deviations from
equilibrium of ionization and from electron-proton temperature
equilibration, the heating due to radioactive decay of unstable
elements, and the back-reaction of accelerated CRs (e.g.
\citealt{2016ApJ...822...22O, 2020arXiv200901789O}). The SNR models
also describe the geometry and density structure of the ambient
medium through which the remnant expands on the base of constraints
from multi-$\lambda$ observations.

This approach also includes the synthesis of both thermal and
non-thermal emission from the models for a qualitative and quantitative
comparison of the model results with specific observations (e.g.
\citealt{2007A&A...470..927O, 2011A&A...526A.129O, 2019NatAs...3..236M}).
This is an essential step to disentangle the effects of CRs acceleration
from other effects as, for instance, the intrinsic ejecta structure
reflecting the asymmetries in the SN explosion and the interaction
of the remnant with the inhomogeneous ISM.

In the following, we discuss a few examples to show how the
above approach can be used to extract various pieces of information
on the acceleration of paticles in SNRs from the comparison of model
results with observations. More specifically, in Sect.~\ref{sec3.1},
we discuss the need to identify the effects of CR acceleration on
the structure of the mixing region in SNRs and to disentangle these
effects from those of the intrinsic structure of the ejecta. In
Sect.~\ref{sec3.2}, we highlight the importance of modeling the
interaction of SNRs with the inhomogeneous ISM (in particular with
atomic/molecular clouds), since the interaction sites provide
excellent opportunities to prove unambiguously CRs acceleration
(being the densities in clouds high enough to allow the $\gamma$-ray
emission to be dominated by protons) and to constrain the energetics
of particle accelerators.  Finally, in Sect.~\ref{sec3.3}, we
consider the process of particle acceleration during the transition
from the phase of SN to that of SNR, a time in which the particles
start to be accelerated and any variations of the acceleration
process in these early times after the SN may be reflected in the
shape of $\gamma$-ray spectra in young and middle-age SNRs
(e.g.~\citealt{2017A&A...605A.110P}).

\subsection{Back-reaction of accelerated CRs and clumping of ejecta}
\label{sec3.1}

\begin{figure}[t]
\centerline{
\includegraphics[width=12cm]{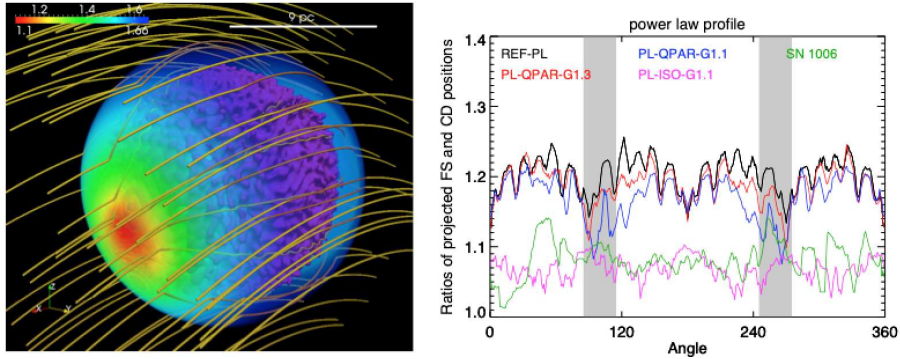}}
\caption{3D rendering of the spatial distribution of the effective
adiabatic index (left panel) and azimuthal profiles of the ratio
between the FS and CD radii when the aspect angle is 90 degrees
(right panel), for models including the back-reaction of accelerated
CRs without the clumping of ejecta. The violet surface tracks the
ejecta material, the yellow lines are sampled magnetic field lines.
Black, red and blue profiles mark the results from the models; the
green line marks the profile derived from observations of SN~1006
(\citealt{2009A&A...501..239M}). The gray areas in the right
panel mark the regions where the acceleration of CRs is the largest.
Figure adapted from \cite{2012ApJ...749..156O}.}
\label{fig1a}
\end{figure}

\begin{figure}[t]
\centerline{
\includegraphics[width=12cm]{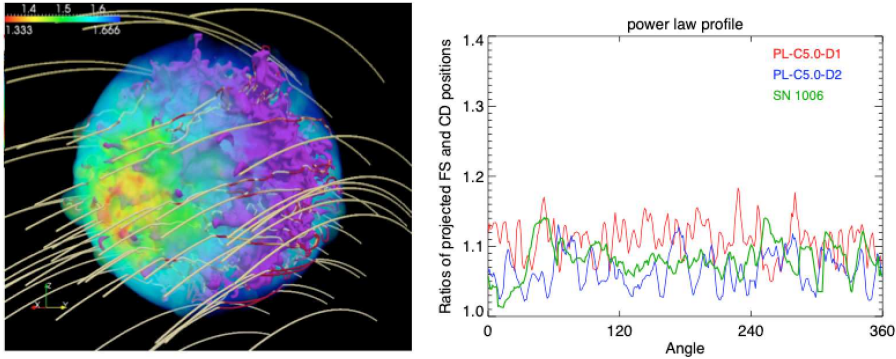}}
\caption{As in Fig.~\ref{fig1a}, but for models including the
clumping of ejecta in addition to the back-reaction of accelerated
CRs. Figure adapted from \cite{2012ApJ...749..156O}.}
\label{fig1b}
\end{figure}

The numerical approach described above was used to investigate the
role played by back-reaction of accelerated CRs and by the intrinsic
structure of the ejecta in determining the ratio of FS to CD radii
(\citealt{2012ApJ...749..156O}). Figures \ref{fig1a} and \ref{fig1b}
show the results of models either with or without an initial
small-scale structure of ejecta, respectively. The models assume
that the effects of the back-reaction of accelerated CRs is modulated
by the obliquity angle and that they are more effective at parallel
shocks where the adiabatic index reaches values as low as 1.1 (in
models without ejecta clumping, thus indicating extreme energy
losses to accelerate the CRs) or 4/3 (in models including the ejecta
clumping).

Figure \ref{fig1a} shows that none of the models without ejecta
clumping and with a dependence on the obliquity angle is able to
reproduce the profile inferred from the observations of SN~1006
(green line). A model without ejecta clumping can reproduce the
observations only if the effects of back-reaction of accelerated
CRs are extreme (with adiabatic index 1.1) at all obliquity angles
(see the purple profile in the upper right panel), a condition which
is not realistic. In models including the ejecta clumping (Fig.~\ref{fig1b}),
clumps and filamentary structures are evident within the remnant,
and fingers of ejecta are close to or protrude beyond the main blast
wave as observed by \cite{2011ApJ...735L..21R} in SN~1006. Indeed,
these models can naturally reproduce the observed azimuthal profile
of the ratio between the FS and CD radii in SN~1006 (see
right panel in Fig.~\ref{fig1b}). This study unambiguously shows that the FS-CD
separation is a probe of the ejecta structure rather than a probe
of the efficiency of CRs acceleration in young SNRs.

\subsection{Interaction of supernova remnant with interstellar clouds}
\label{sec3.2}

The interaction of SNRs with atomic or molecular clouds is quite
common (a few examples are IC 443 and W48b) and it is important
for the acceleration of CRs. For instance, MHD simulations of the
shock-cloud interaction revealed a crucial role played by the
magnetic field (\citealt{2016MNRAS.456.2343P, 2018MNRAS.479.4253P}):
the tangential magnetic field component prevents a large compression
of post-adiabatic shocks visible in the pure HD simulations. This
effect may lower the $\gamma$-ray flux arising from the shock-cloud
interaction region. IC 443 is an example of remnant interacting
with clouds; it is a strong $\gamma$-ray source and one of the few
remnants providing evidence for hadronic CRs acceleration. Observations
show a diffuse $\gamma$-ray emission coincident with the FS of IC
443; a clear enhancement of $\gamma$-ray emission is evident in the
northeast (NE) rim where the remnant is interacting with a molecular
cloud.

SN~1006 is another remnant showing evidence of interaction with an
atomic cloud. This SNR is an interesting target to study particle
acceleration because of the bilateral morphology of its non-thermal
emission which reflects efficient particle acceleration in the NE
and southwest (SW) limbs bright in the radio, X-ray and $\gamma$-ray
bands. XMM observations of this remnant revealed a sharp
indentation in the SW limb and several signatures for a shock-cloud
interaction (\citealt{2014ApJ...782L..33M}) in a region corresponding
to the position of an H\,I cloud. A clear proof of the shock-cloud
interaction is provided by the azimuthal profile of the cutoff
energy which has a minimum in coincidence with the indentation.

\begin{figure}[t]
\centerline{
\includegraphics[width=10cm]{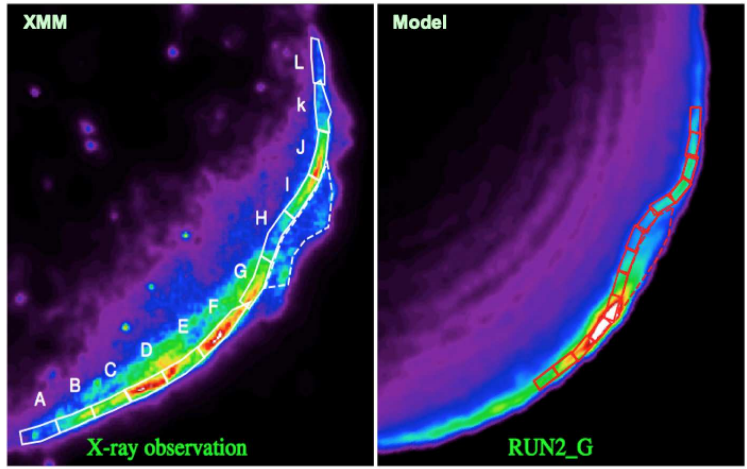}}
\caption{XMM-Newton (on the left) and modeled (on the
right) images of the SW limb of SN~1006 in the $[2, 4.5]$~keV band.
The images show the regions selected for the spectral analysis of
the rim. Figure adapted from \cite{2016A&A...593A..26M}.}
\label{fig2a}
\end{figure}

\begin{figure}[t]
\centerline{
\includegraphics[width=7.5cm]{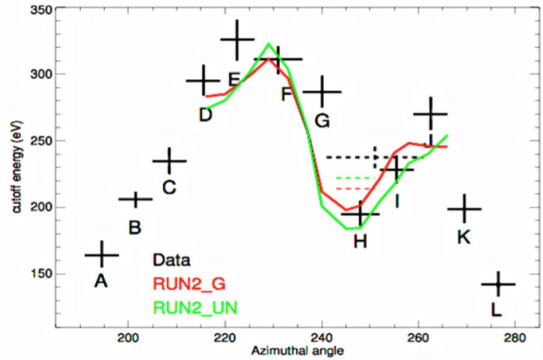}}
\caption{
Azimuthal variations of the synchrotron cutoff
energy. The black crosses show the best fit values obtained from
the analysis of the X-ray spectra extracted from the regions marked
in Fig.~\ref{fig2a}. The green/red curves show the values derived
from two MHD models. Figure adapted from \cite{2016A&A...593A..26M}.}
\label{fig2b}
\end{figure}

The shock-cloud interaction was investigated through 3D MHD
simulations, including the back-reaction of particle acceleration,
and describing the interaction of the remnant with the H\,I cloud
(\citealt{2016A&A...593A..26M}). Fig.~\ref{fig2a} shows the XMM
observation of the SW limb and the synthetic non-thermal X-ray image
derived from the best-fit model. The model roughly describe the
distribution of surface brightness and the indentation observed in
XMM observations. Also, the model is able to reproduce the azimuthal
profile of cutoff energy inferred from observations, and in particular
the drop in energy in coincidence with the indentation (see
Fig.~\ref{fig2b}). The synthesis of the hadronic and leptonic
emission in the $\gamma$-ray band from the models and the comparison
of the model results with observations collected by HESS and FERMI
allowed to constrain the total hadronic energy to $\approx 5\times
10^{49}$~erg, a value confirmed later by \cite{2017ApJ...851..100C}
who found that this hadronic energy is conceivable to explain the
$\gamma$-ray emission in the SW region of SN~1006.

\subsection{Particle acceleration in very young SNRs: the case of
SN 1987A}
\label{sec3.3}

Among other things, very young SNRs offer the opportunity to study
the process of particle acceleration during the transition from the
phase of SN to that of SNR. One of the best case to study is certainly
SN 1987A. The SN was observed in February 1987 and, since then, was
continuously monitored at all wavelength bands. About three years
later the blast wave started to interact with a dense circumstellar
nebula consisting of a dense equatorial ring and an extended H\,II
region.

\begin{figure}[t]
\centerline{
\includegraphics[width=13cm]{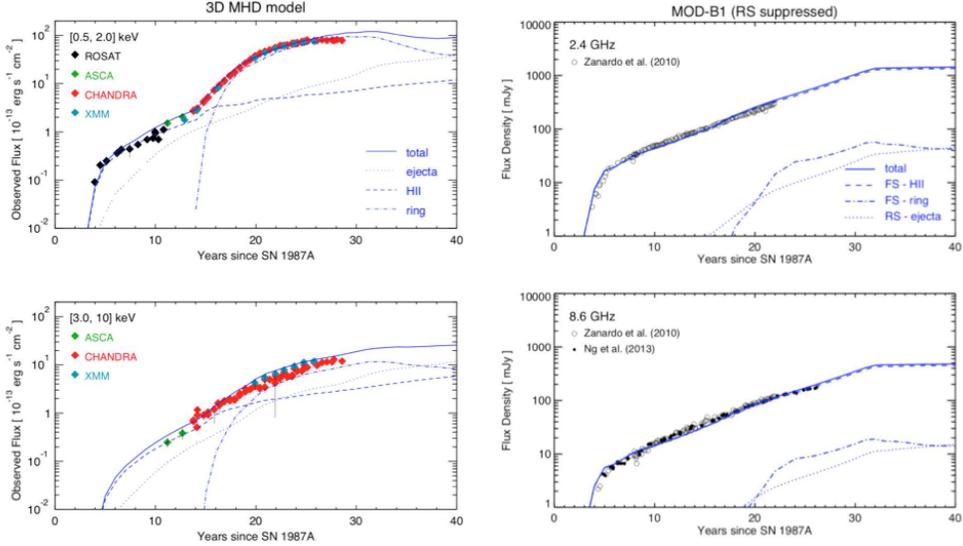}}
\caption{Observed (symbols) and synthetic (lines) X-ray lightcurves
in the $[0.5, 2]$~keV (upper left panel) and $[3, 10]$~keV (lower
left panel) bands and radio flux densities at 2.4 GHz (upper right
panel; \citealt{2010ApJ...710.1515Z}) and 8.6 GHz (lower right
panel; \citealt{2010ApJ...710.1515Z, 2013ApJ...777..131N}). Solid
lines show the synthetic lightcurves; dotted, dashed, and dot-dashed
lines indicate the contribution to emission from the shocked ejecta,
the shocked plasma from the H II region, and the shocked plasma
from the ring, respectively. The model assumes that the radio
emission from the RS is reduced by two orders of magnitude. Figure
adapted from \cite{2019A&A...622A..73O}.}
\label{fig3}
\end{figure}

The model approach described above was adopted to investigate the
evolution of SN 1987A from the immediate aftermath of the core-collapse
of the progenitor star to the subsequent interaction of its remnant
with the nebula. The model includes the synthesis of thermal X-ray
emission and non-thermal radio emission to compare the model results
with observations of SN 1987A. The observed (and modeled) X-ray (in
both the soft and hard bands) and radio lightcurves show that, few
years after the SN, both X-ray and radio fluxes suddenly increased
after the impact of the blast wave with the H\,II region (see
Fig.~\ref{fig3}). In the subsequent years the X-ray and radio
fluxes have continued to rise together with almost constant slope.
About 14 years after the SN the soft X-ray flux has steepened still
further due to the interaction with the dense equatorial ring,
contrary to the hard X-ray and radio lightcurves which continued
to rise with almost constant slope.

Thanks to the model it was possible to explain the origin of the
X-ray and radio emission. In fact, the hard X-ray emission is almost
insensitive to the interaction of the blast wave with the ring
because the latter is very dense and the shocked ring material emits
mainly in the soft X-ray band. Conversely, most of the contribution
to the hard X-ray emission originates from shocked material from
the H\,II region. This determines the sudden rise of the soft X-ray
emission during the interaction with the ring and an almost constant
slope of the hard X-ray emission (\citealt{2015ApJ...810..168O}).
The 3D MHD model of SN 1987A also provided a natural explanation
for the origin of the observed trend of the radio emission (similar
to that of the hard X-ray emission). In particular, the model can
reproduce the observed radio lightcurves if the radio flux originating
from the RS is partially suppressed (see Fig.~\ref{fig3};
\citealt{2019A&A...622A..73O}). As a result, the radio emission
originates mostly from the FS traveling through the H\,II region
(the contribution from the FS traveling through the ring is at least
an order of magnitude lower) and this explains why the radio emission
is insensitive to the interaction of the blast with the ring. The
analysis of possible mechanisms for the suppression of emission
from the RS has shown that synchrotron self-absorption and free–free
absorption have negligible effects on the emission. Thus the model
suggests that the emission from the RS at radio frequencies might
be limited by highly magnetized ejecta (\citealt{2019A&A...622A..73O}).

\section{Summary and conclusions}

In this contribution, we discussed how models describing the evolution
of SNRs can include the macroscopic effects of particle acceleration
at shock fronts, discussing possible approaches and methodologies.
These models do not describe in detail the physics of particle
acceleration as PIC/hybrid simulations based on first principles
do (e.g. \citealt{2014ApJ...783...91C, 2014ApJ...794...46C}). On
the other side, they allow to explore large dynamical ranges,
describing the evolution of SNRs (since the SN explosion) and their
interaction with the inhomogeneous ISM (e.g.  \citealt{2015ApJ...810..168O,
2016ApJ...822...22O}).

Current limitations of these models include an accurate description
of the back-reaction of accelerated particles which requires input
from first principle simulations. In recent years, many authors
have stressed that identifying the signature of CRs acceleration
requires to link SNRs to their parent SN explosions and progenitor
stars. In fact, establishing this connection is important to describe
accurately the structure of ejecta which may be characterized by
anisotropies developed soon after the core-collapse and the structure
of the circumstellar medium sculpted by the winds of the progenitor
star. This is an essential step to disentangle the effects of CRs
acceleration from other effects determining the structure and
morphology of SNRs in view of an accurate comparison of model results
with multi-$\lambda$ observations of SNRs.

\bigskip
{\bf Acknowledgements.}
We thank the anonymous referee for useful suggestions that have
allowed us to improve the paper. We acknowledge financial contribution
from the INAF mainstream program and from the agreement ASI-INAF
n.2017-14-H.O.




\bibliographystyle{elsarticle-harv}
\bibliography{references}

%
%
%
%

\end{document}